\documentstyle[11pt,bezier,twoside]{article}
\def\greaterthansquiggle{\raise.3ex\hbox{$>$\kern-.75em\lower1ex\hbox{$\sim$}}}
\def\lessthansquiggle{\raise.3ex\hbox{$<$\kern-.75em\lower1ex\hbox{$\sim$}}}
\newcommand{\beq}{\begin{equation}}
\newcommand{\eeq}{\end{equation}}
\newcommand{\beqa}{\begin{eqnarray}}
\newcommand{\eeqa}{\end{eqnarray}}
\newcommand{\beqan}{\begin{eqnarray*}}
\newcommand{\eeqan}{\end{eqnarray*}}
\newcommand{\ba}{\begin{array}}
\newcommand{\ea}{\end{array}}
\newcommand{\no}{\nonumber}

\newcommand{\ra}{\rightarrow}

\newcommand{\cL}{{\cal L}}

\newcommand{\cO}{{\cal O}}

\newcommand{\dfrac}{\displaystyle \frac}

\def\nz{\ifmmode {I\hskip -3pt N} \else {\hbox {$I\hskip -3pt N$}}\fi}
\def\zz{\ifmmode {Z\hskip -4.8pt Z} \else
       {\hbox {$Z\hskip -4.8pt Z$}}\fi}
\def\qz{\ifmmode {Q\hskip -5.0pt\vrule height6.0pt depth 0pt
       \hskip 6pt} \else {\hbox
       {$Q\hskip -5.0pt\vrule height6.0pt depth 0pt\hskip 6pt$}}\fi}
\def\rz{\ifmmode {I\hskip -3pt R} \else {\hbox {$I\hskip -3pt R$}}\fi}
\def\cz{\ifmmode {C\hskip -4.8pt\vrule height5.8pt\hskip 6.3pt} \else
       {\hbox {$C\hskip -4.8pt\vrule height5.8pt\hskip 6.3pt$}}\fi}

\def\au{{\setbox0=\hbox{\lower1.36775ex%
\hbox{''}\kern-.05em}\dp0=.36775ex\hskip0pt\box0}}
\def\ao{{}\kern-.10em\hbox{``}}

\normalsize
\begin{document}
\bibliographystyle{plain}
\begin{titlepage}
\begin{flushright}
UWThPh-1995-14\\
June 6, 1995
\end{flushright}
\vspace*{3cm}
\begin{center}
{\Large \bf
The decay $\!\mbox{\boldmath $\rho^0 \ra \pi^+ \pi^- \gamma$}$ and the
\\[7pt]
chiral invariant interactions of vector mesons*}\\[50pt]
K. Huber and H. Neufeld \\
Institut f\"ur Theoretische Physik der Universit\"at Wien\\
Boltzmanngasse 5, A-1090 Wien, Austria \\

\vfill
{\bf Abstract} \\
\end{center}
\noindent
Using the close relationship between the low--energy constants of
chiral perturbation theory and the chiral invariant interactions of
the vector meson resonances with the pseudoscalar mesons, we
investigate the process $\rho^0 \ra \pi^+ \pi^- \gamma$. Compared
with the contribution from the pure bremsstrahlung mechanism, we find
an enhancement of the decay rate near the endpoint of the photon
energy spectrum. Such a particular shape of the differential decay rate
has indeed been observed experimentally and turns out to be an important
confirmation of the theoretical concept of chiral vector meson dominance.
\vfill
\noindent * Work supported in part by Fonds zur F\"orderung der
wissenschaftlichen Forschung (Austria), Project No. P09505--PHY and by HCM,
EEC--Contract No. CHRX--CT920026 (EURODA$\Phi$NE).
\end{titlepage}

\section{Introduction}
\label{sec: Introduction}
\renewcommand{\theequation}{\arabic{section}.\arabic{equation}}
\setcounter{equation}{0}
At low energies, the interactions of the pseudoscalar mesons are
described by chiral perturbation theory ($\chi$PT)
\cite{Weinberg,GL1,GL2}. This effective low--energy theory is
mathematically equivalent \cite{Leutwyler} to QCD, the underlying
fundamental theory. The lagrangian of $\chi$PT contains all terms
compatible with the symmetries of QCD. The associated coupling
constants can be interpreted as describing the influence of all
degrees of freedom not explicitly contained in the chiral lagrangian.

It has been shown \cite{EGPR,EGLPR,DRV} that the coupling constants
which occur at $\cO(p^4)$ in the low--energy expansion are dominated
by the low--lying vector, axial--vector, scalar and pseudoscalar
resonances. In particular, chiral symmetry was found to confirm the
phenomenologically successful concept of vector meson dominance
\cite{Sakurai}.

In this letter we shall employ this connexion between the chiral
invariant vector meson interactions and the low--energy constants of
$\chi$PT. It is possible to obtain rather precise results for the
decay spectrum of $\rho^0 \ra \pi^+ \pi^- \gamma$ which can be
confronted with the present experimental data
\cite{Vasserman,Dolinsky}.

\section{The low--energy limit of QCD}
\label{sec: The low--energy limit of QCD}
\renewcommand{\theequation}{\arabic{section}.\arabic{equation}}
\setcounter{equation}{0}

The strong, electromagnetic and semileptonic interactions of
pseudoscalar mesons are described by an effective chiral lagrangian
$\cL_{\rm eff}$. It consists of a string of terms
\beq
\cL_{\rm eff} = \cL_2 + \cL_4 + \cL_6 + ... \enspace ,
\eeq
organized in powers of momenta and meson masses, respectively. The
lowest order term $\cL_2$ is the nonlinear sigma model lagrangian in
the presence of external fields\footnote{Our notation is the same as
in refs. \cite{EGPR,EGLPR}.}:
\beq
\cL_2 = \frac{F^2}{4} \langle u_\mu u^\mu + \chi_+\rangle, \label{L2}
\eeq

The generating functional $Z[v,a,s,p]$ for the Green functions of vector,
axial--vector, scalar and pseudoscalar quark currents is
given by the expansion of the effective meson field theory in the
number of loops,
\beq
Z = Z_2 + Z_4 + Z_6 + ... \enspace .
\eeq
The leading term coincides with the classical action associated with
$\cL_2$.

At next to leading order $p^4$ the generating functional consists of
the following terms: one--loop graphs generated by the vertices of
$\cL_2$, tree graphs involving one vertex from $\cL_4$ and finally a
contribution $Z_{W\!ZW}$ to account for the chiral anomaly. The chiral
invariant lagrangian $\cL_4$ of $\cO(p^4)$ is given by \cite{GL2}
\beqa
\cL_4 &=& L_1 \; \langle u_\mu u^\mu \rangle^2 + L_2 \; \langle u_\mu
u^\nu\rangle \; \langle u^\mu u_\nu\rangle
+ L_3 \; \langle u_\mu u^\mu u_\nu u^\nu\rangle +
L_4 \; \langle u_\mu u^\mu\rangle \; \langle \chi_+\rangle \no \\
&& \mbox{} + L_5 \; \langle u_\mu u^\mu \chi_+\rangle +
L_6 \; \langle \chi_+\rangle^2 + L_7 \; \langle \chi_-\rangle^2
+ \frac{1}{4} (2L_8 + L_{12}) \langle \chi_+^2\rangle +
\frac{1}{4} (2L_8 - L_{12}) \langle \chi_-^2\rangle \no \\
&& \mbox{} - iL_9 \; \langle f_+^{\mu\nu} u_\mu u_\nu\rangle +
\frac{1}{4} (L_{10} + 2L_{11}) \langle f_{+\mu\nu} f_+^{\mu\nu}\rangle
- \frac{1}{4} (L_{10} - 2L_{11})
\langle f_{-\mu\nu} f_-^{\mu\nu}\rangle. \label{L4}
\eeqa
The twelve low--energy couplings $L_1,\ldots,L_{12}$ arising here are
divergent (except $L_3$, $L_7$). They absorb the divergences of the
one--loop graphs via the renormalization
\beqa
L_i &=& L_i^r(\mu) + \Gamma_i \Lambda(\mu) , \no \\
\Lambda(\mu) &=& \frac{\mu^{d-4}}{(4\pi)^2} \left\{ \frac{1}{d-4} -
\frac{1}{2} [\ln (4\pi) + \Gamma'(1) + 1]\right\}, \label{renorm}
\eeqa
in the dimensional regularization scheme.

The meson decay constant $F \simeq F_{\pi} = 92.4 \mbox{ MeV}$
\cite{PDG}, the
vacuum condensate parameter $B_0$ (contained in $\chi_+$), together
with $L^r_1,\ldots,L^r_{10}$,
determine the low--energy behaviour of pseudoscalar meson interactions
to $\cO(p^4)$. $L^r_{11}$ and $L^r_{12}$ are contact terms which are
not directly accessible to experiment. The coupling constants
$L_1^r,\ldots,L^r_{10}$ have been determined from experimental input
and by using large--$N_c$ arguments \cite{GL2}. An updated list of
their numerical values can be found in ref. \cite{DAFNE2}.

\section{Chiral couplings of vector resonances}
\label{sec: Chiral couplings of vector resonances}
\renewcommand{\theequation}{\arabic{section}.\arabic{equation}}
\setcounter{equation}{0}

To investigate the contribution of vector meson exchange to the
$L_i^r$, one has to include spin--1 fields in $\cL_{\rm eff}$. Following
ref. \cite{EGPR}, we describe the vector meson octet $(\rho, K^*,
\omega_8)$ in terms of a $3 \times 3$ matrix valued antisymmetric
tensor field $V_{\mu \nu}$.
The kinetic term of the vector meson lagrangian takes the form
\beq
\cL_{\rm kin}(V) = - \frac{1}{2} \; \left\langle \nabla^\lambda
V_{\lambda\mu} \nabla_\nu V^{\nu\mu} -
\frac{M_V^2}{2} V_{\mu\nu} V^{\mu\nu}\right\rangle. \label{Lkin}
\eeq
To lowest order $p^2$ in the chiral expansion, the most general
interaction of a single vector meson with the pseudoscalar bosons
$(\pi, K, \eta)$ is described by the lagrangian \cite{EGPR}
\beq
\cL_2(V) =
\frac{F_V}{2\sqrt{2}}\; \langle V_{\mu\nu} f_+^{\mu\nu} \rangle +
\frac{i G_V}{\sqrt{2}} \; \langle V_{\mu\nu} u^\mu u^\nu \rangle,
\label{L2V}
\eeq
where the two real coupling constants $F_V$ and $G_V$ are not further
restricted by chiral symmetry. Their numerical values can be
determined from the decay rates for $\rho^0 \ra e^+ e^-$ and $\rho
\ra \pi \pi$. From (\ref{L2V}) one obtains
\beq
\Gamma(\rho^0 \ra e^+ e^-) = \frac{4 \pi \alpha^2 F_V^2}{3 M_{\rho}}
(1 + \frac{2 m_e^2}{M_{\rho}^2}) (1 - \frac{4
m_e^2}{M_{\rho}^2})^{1/2}, \label{RHOEP}
\eeq
and
\beq
\Gamma(\rho \ra \pi \pi) = \frac{G_V^2 M_{\rho}^3}{48 \pi F_{\pi}^4}
(1 - \frac{4 M_{\pi}^2}{M_{\rho}^2})^{3/2}, \label{RHOPIPI}
\eeq
respectively. Comparison with the experimental value
$\Gamma(\rho^0 \ra e^+ e^-) = (6.7 \pm 0.3) \mbox{ keV}$ \cite{PDG} yields
\beq
\left| F_V \right| \simeq 153 \mbox{ MeV}, \label{FV}
\eeq
while $\Gamma(\rho \ra \pi \pi) = (151.2 \pm 1.2) \mbox{ MeV}$
\cite{PDG} implies
\beq
\left| G_V \right| \simeq 67 \mbox{ MeV}. \label{GV}
\eeq
Higher order chiral corrections may reduce this value to
$\left| G_V \right| \simeq 53 \mbox{ MeV}$ \cite{EGPR}.

To order $p^4$ in the chiral expansion, vector meson exchange induces
a local lagrangian of the type (\ref{L4}) with \cite{EGPR}
\beq
\ba{l}
L_1^V = \dfrac{G_V^2}{8 M_V^2}, \quad L_2^V = 2 L_1^V, \quad
L_3^V = -6 L_1^V, \no \\[10pt]
L_9^V = \dfrac{F_V G_V}{2 M_V^2}, \quad
L_{10}^V = - \dfrac{F_V^2}{4 M_V^2},  \no \\[10pt]
L_4^V = L_5^V = L_6^V = L_7^V = L_8^V = 0.
\ea
\label{LiV}
\eeq
With $M_V = M_{\rho}$, the information on $\left| F_V \right|$ from
$\rho^0 \ra e^+ e^-$ and on $\left| G_V \right|$ from $\rho \ra \pi
\pi$, the contributions (\ref{LiV}) (including $L_{10}^A$ from
axial--vector exchange) were found \cite{EGPR} to saturate the
phenomenological values of the renormalized coupling constants
$L_i^r(M_{\rho})$ for $i = 1,2,3,9,10$.
In addition, the coupling constants with vanishing contributions from
the vector (or axial--vector) mesons are saturated by scalar and
pseudoscalar exchange, respectively \cite{EGPR}.
These findings can be summarized by means of the following
catch--words \cite{Ecker}:
\begin{itemize}
\item {\bf Chiral duality:} The $L_i^r(M_{\rho})$ are practically
saturated by resonance exchange.
\item {\bf Chiral vector meson dominance:} Whenever spin--1
resonances can contribute at all $(i =1,2,3,9,10)$, the
$L_i^r(M_{\rho})$ are almost completely dominated by vector and
axial--vector exchange.
\end{itemize}
Furthermore, with additional QCD--inspired assumptions, the vector
and axial--vector contributions to the $L_i$ can be expressed in
terms of $F_{\pi}$ and $M_V \simeq M_{\rho}$ \cite{EGLPR},
\beq
8 L_1^V = 4 L_2^V = -\frac{4}{3} L_3^V = L_9^V = -\frac{4}{3}
L_{10}^{V+A} = \frac{F_{\pi}^2}{2 M_V^2}, \label{QCD}
\eeq
in very good agreement with the phenomenological values of the
$L_i^r(M_{\rho})$.

\section{The decay $\rho^0 \ra \pi^+ \pi^- \gamma$}
\label{sec: The decay}
\renewcommand{\theequation}{\arabic{section}.\arabic{equation}}
\setcounter{equation}{0}

The observed rates for $\rho^0 \ra e^+ e^-$ and $\rho \ra \pi \pi$
permit only the determination of the absolute values of the coupling
constants $F_V, G_V$. In the case of $\rho^0 \ra \pi^+ \pi^- \gamma$
there is also a contribution from an interference term between $F_V$
and $G_V$ which provides us with the additional experimental
information about the sign of the product $F_V G_V$. As we have seen
in the last section, chiral vector meson dominance relates this
quantity to the low--energy constant $L_9^r$:
\beq
L_9^r(M_{\rho}) \simeq L_9^V = \frac{F_V G_V}{2 M_V^2}. \label{L9}
\eeq
On the other hand, a very precise numerical value \cite{DAFNE2} for
$L_9^r(M_{\rho})$ has been extracted from the pion charge radius:
\beq
L_9^r(M_{\rho}) = (6.9 \pm 0.7) \cdot 10^{-3}.
\eeq
Together with (\ref{L9}), this clearly implies $F_V G_V > 0$.
The same conclusion can, of course, also be drawn from (\ref{QCD}).

Taking the relevant vertices from (\ref{L2V}) and (\ref{L2}), the
decay rate of $\rho^0 \ra \pi^+ \pi^- \gamma$ can easily be
calculated\footnote{An older theoretical treatment can be found
in Ref. \cite{Singer}.}. The photon spectrum is given by
\beqa
\frac{d\Gamma}{dE_{\gamma}} &=& \frac{e^2 M_{\rho}^2}{96 \pi^3 F_{\pi}^4 z}
\left\{ \left[F_V^2 z^4 +2 F_V G_V z^2 (1 - 2 z^2)
+ G_V^2 \left(4 z^4 + (2 z - 1) (1 - 4 r_{\pi}^2)
\right)\right] w(z) \right. \no \\
&& \left.  \mbox{} +  \left[- 4 F_V G_V r_{\pi}^2 z^2
+ G_V^2 \left(1 - 2 z +
2 r_{\pi}^2 (4 z^2 + 4 z -3 + 4 r_{\pi}^2)\right)\right]
{\rm ln}\frac{1 + w(z)}{1 - w(z)} \right\},
\eeqa
with
\beq
z = E_{\gamma}/M_{\rho}, \quad w(z) = \sqrt{1 - \frac{4 r_{\pi}^2}{1 -
2 z}}, \quad  r_{\pi} = M_{\pi}/M_{\rho},
\eeq
where $E_{\gamma}$ is the photon energy in the rest frame of the $\rho^0$.

In our numerical analysis, the input parameters $F_V, G_V$ have been
chosen in such a way that the central values of the present
experimental data for $\Gamma(\rho^0 \ra e^+ e^-)$ and
$\Gamma(\rho \ra \pi \pi)$ (see sect. \ref{sec: Chiral couplings of
vector resonances}) are reproduced by (\ref{RHOEP}) and
(\ref{RHOPIPI}), respectively. Our results for $\rho^0 \ra \pi^+
\pi^- \gamma$ are displayed in fig.~1. The thick solid curve shows
the theoretically expected differential decay rate with $F_V G_V >
0$. For comparison, the dotted line gives the contribution of pure
bremsstrahlung ($F_V = 0$), while the thin solid curve corresponds to
$F_V G_V < 0$.
\begin{figure}
\setlength{\unitlength}{0.3mm}
\begin{picture}(350,300)(-90,-30)
\put(0,0){\line(1,0){350}}
\put(0,270){\line(1,0){350}}
\put(0,0){\line(0,1){270}}
\put(350,0){\line(0,1){270}}
\multiput(50,0)(50,0){6}{\line(0,1){10}}
\multiput(50,270)(50,0){6}{\line(0,-1){10}}
\multiput(0,50)(0,50){5}{\line(1,0){10}}
\multiput(0,10)(0,10){26}{\line(1,0){5}}
\multiput(350,50)(0,50){5}{\line(-1,0){10}}
\multiput(350,10)(0,10){26}{\line(-1,0){5}}
\put(0,-20){\makebox(0,0){0}}
\put(50,-20){\makebox(0,0){50}}
\put(100,-20){\makebox(0,0){100}}
\put(150,-20){\makebox(0,0){150}}
\put(200,-20){\makebox(0,0){200}}
\put(250,-20){\makebox(0,0){250}}
\put(300,-20){\makebox(0,0){300}}
\put(350,-20){\makebox(0,0){350}}
\put(175,-45){\makebox(0,0){$E_{\gamma} \, \mbox{(MeV)}$}}
\put(-25,100){\makebox(0,0){$1 \cdot 10^{-2}$}}
\put(-25,200){\makebox(0,0){$2 \cdot 10^{-2}$}}
\put(250,200){\makebox(0,0){$d \, \Gamma (\rho^0 \rightarrow \pi^+
\pi^- \gamma) / d E_{\gamma}$ }}
\thicklines
\bezier{112}(50,269.7)(55,240)(60,215.4)
\bezier{79}(60,215.4)(65,194.6)(70,177)
\bezier{61}(70,177)(75,161.6)(80,148.3)
\bezier{48}(80,148.3)(85,136.6)(90,126.2)
\bezier{39}(90,126.2)(95,117)(100,108.8)
\bezier{34}(100,108.8)(105,101.3)(110,94.6)
\bezier{33}(110,94.6)(115,88.5)(120,83)
\bezier{28}(120,83)(125,77.9)(130,73.3)
\bezier{26}(130,73.3)(135,69)(140,65.2)
\bezier{24}(140,65.2)(145,61.5)(150,58.2)
\bezier{23}(150,58.2)(155,55.1)(160,52.2)
\bezier{22}(160,52.2)(165,49.5)(170,47)
\bezier{21}(170,47)(175,44.7)(180,42.6)
\bezier{21}(180,42.6)(185,40.5)(190,38.6)
\bezier{21}(190,38.6)(195,36.8)(200,35.2)
\bezier{21}(200,35.2)(205,33.6)(210,32.1)
\bezier{21}(210,32.1)(215,30.7)(220,29.5)
\bezier{20}(220,29.5)(225,28.2)(230,27.1)
\bezier{20}(230,27.1)(235,25.9)(240,24.9)
\bezier{20}(240,24.9)(245,23.9)(250,23)
\bezier{20}(250,23)(255,22.1)(260,21.2)
\bezier{20}(260,21.2)(265,20.3)(270,19.6)
\bezier{20}(270,19.6)(275,18.8)(280,18)
\bezier{20}(280,18)(285,17.2)(290,16.5)
\bezier{20}(290,16.5)(295,15.7)(300,14.8)
\bezier{20}(300,14.8)(305,13.9)(310,12.9)
\bezier{21}(310,12.9)(315,11.8)(320,10.5)
\bezier{22}(320,10.5)(325,8.7)(330,6.2)
\bezier{5}(330,6.2)(331,5.5)(332,4.6)
\bezier{7}(332,4.6)(333,3.5)(334,1.8)
\bezier{3}(334,1.8)(334.3,1)(334.3,0)
\bezier{11}(50,267.3)(55,237.4)(60,212.6)
\bezier{8}(60,212.6)(65,191.6)(70,173.7)
\bezier{6}(70,173.7)(75,158.1)(80,144.6)
\bezier{5}(80,144.6)(85,132.6)(90,122.1)
\bezier{4}(90,122.1)(95,112.6)(100,104.1)
\bezier{4}(100,104.1)(105,96.5)(110,89.6)
\bezier{3}(110,89.6)(115,83.2)(120,77.5)
\bezier{3}(120,77.5)(125,72.3)(130,67.4)
\bezier{3}(130,67.4)(135,62.9)(140,58.8)
\bezier{2}(140,58.8)(145,55.0)(150,51.5)
\bezier{2}(150,51.5)(155,48.1)(160,45.1)
\bezier{2}(160,45.1)(165,42.2)(170,39.6)
\bezier{2}(170,39.6)(175,37.0)(180,34.7)
\bezier{2}(180,34.7)(185,32.5)(190,30.5)
\bezier{2}(190,30.5)(195,28.5)(200,26.7)
\bezier{2}(200,26.7)(205,25.0)(210,23.4)
\bezier{2}(210,23.4)(215,21.8)(220,20.4)
\bezier{2}(220,20.4)(225,19.1)(230,17.8)
\bezier{2}(230,17.8)(235,16.6)(240,15.4)
\bezier{2}(240,15.4)(245,14.4)(250,13.4)
\bezier{2}(250,13.4)(255,12.4)(260,11.5)
\bezier{2}(260,11.5)(265,10.6)(270,9.8)
\bezier{2}(270,9.8)(275,9.1)(280,8.3)
\bezier{2}(280,8.3)(285,7.6)(290,7.0)
\bezier{2}(290,7.0)(295,6.4)(300,5.8)
\bezier{2}(300,5.8)(305,5.2)(310,4.6)
\bezier{2}(310,4.6)(315,4.0)(320,3.4)
\bezier{2}(320,3.4)(325,2.7)(330,1.8)
\bezier{1}(330,1.8)(332,1.3)(334.3,0)
\thinlines
\bezier{112}(50,265.0)(55,234.9)(60,209.8)
\bezier{79}(60,209.8)(65,188.6)(70,170.5)
\bezier{61}(70,170.5)(75,154.7)(80,141.0)
\bezier{48}(80,141.0)(85,128.8)(90,118.0)
\bezier{39}(90,118.0)(95,108.4)(100,99.7)
\bezier{34}(100,99.7)(105,91.9)(110,84.8)
\bezier{33}(110,84.8)(115,78.3)(120,72.4)
\bezier{28}(120,72.4)(125,67.0)(130,62.0)
\bezier{26}(130,62.0)(135,57.4)(140,53.1)
\bezier{24}(140,53.1)(145,49.2)(150,45.5)
\bezier{23}(150,45.1)(155,42.1)(160,38.9)
\bezier{22}(160,38.9)(165,36.0)(170,33.2)
\bezier{21}(170,33.2)(175,30.6)(180,28.2)
\bezier{21}(180,28.2)(185,25.9)(190,23.8)
\bezier{21}(190,23.8)(195,21.8)(200,20.0)
\bezier{21}(200,20.0)(205,18.2)(210,16.6)
\bezier{21}(210,16.6)(215,15.1)(220,13.7)
\bezier{20}(220,13.7)(225,12.3)(230,11.1)
\bezier{20}(230,11.1)(235,9.9)(240,8.8)
\bezier{20}(240,8.8)(245,7.8)(250,6.9)
\bezier{20}(250,6.9)(255,6.0)(260,5.2)
\bezier{20}(260,5.2)(265,4.5)(270,3.9)
\bezier{20}(270,3.9)(275,3.3)(280,2.7)
\bezier{20}(280,2.7)(285,2.2)(290,1.8)
\bezier{20}(290,1.8)(295,1.4)(300,1.07)
\bezier{20}(300,1.07)(305,0.79)(310,0.56)
\bezier{21}(310,0.56)(315,0.37)(320,0.23)
\bezier{22}(320,0.23)(325,0.12)(330,0.05)
\bezier{5}(330,0.05)(331,0.04)(332,0.03)
\bezier{7}(332,0.03)(333,0.02)(334,0.01)
\bezier{3}(334,0.01)(334.3,0)(334.3,0)
\end{picture}
\vspace{1cm}
\caption{Photon spectrum of $\rho^0 \ra \pi^+ \pi^- \gamma$ for $F_V
G_V > 0$ (thick solid curve), $F_V = 0$ (dotted curve) and $F_V G_V <
0$ (thin solid curve).}
\end{figure}
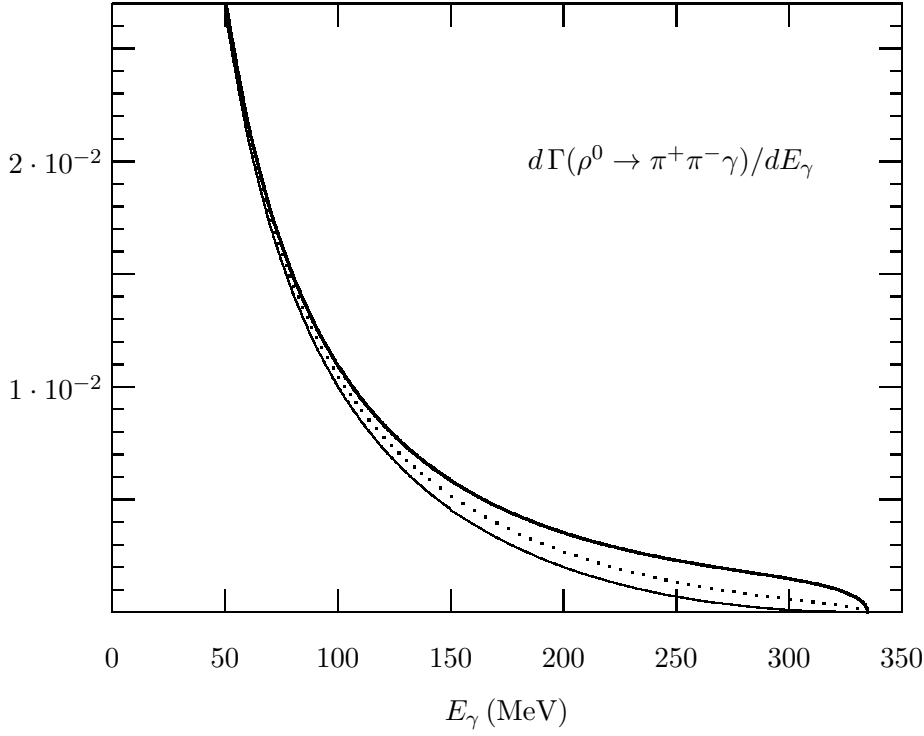

For small photon energies, the decay rate is, of course, dominated by
bremsstrahlung. However, at the upper end of the photon spectrum, the
solution favoured by chiral vector meson dominance shows a sizable
enhancement compared to bremsstrahlung alone.

Experimentally, the decay rate of $\rho^0 \ra \pi^+ \pi^- \gamma$ has
been determined at the $e^+ e^-$ storage ring VEPP--2M at Novosibirsk
\cite{Vasserman,Dolinsky}. Fig.~16 of ref. \cite{Dolinsky} shows the
observed photon spectrum in the energy range $E_{\gamma} > 50 \mbox{
MeV}$. The measured values are rather close to the rate expected from
pure bremsstrahlung, except for the last bin ($E_{\gamma} > 300 \mbox{
MeV}$) where a significant enhancement (three standard deviations)
has been observed.

These experimental results are in perfect agreement with our
theoretical expectations. Not only the presence of the $F_V$ coupling
in $\rho^0 \ra \pi^+ \pi^- \gamma$ has been detected but also the
positive sign of $F_V G_V$ has been clearly established. Note that
$F_V G_V < 0$ would have implied a practically vanishing contribution
to the decay rate for $E_{\gamma} > 300 \mbox{ MeV}$.

Let us finally compare the theoretical and the experimental branching
ratio. For $F_V G_V > 0$ we find
\beq
BR(\rho^0 \ra \pi^+ \pi^- \gamma) = 1.1 \cdot 10^{-2} \quad \mbox{
for} \enspace E_{\gamma} > 50 \mbox{ MeV},
\eeq
while the measured value is given by \cite{Dolinsky}
\beq
BR(\rho^0 \ra \pi^+ \pi^- \gamma) = (0.99 \pm 0.04 \pm 0.15) \cdot
10^{-2} \quad \mbox{ for} \enspace E_{\gamma} > 50 \mbox{ MeV}.
\eeq

\section{Summary}
\label{sec: Summary}
\renewcommand{\theequation}{\arabic{section}.\arabic{equation}}
\setcounter{equation}{0}

The present experimental data for the decay $\rho^0 \ra \pi^+ \pi^-
\gamma$ can be consistently described by the lowest--order vector
meson lagrangian (\ref{L2V}) with two coupling constants $F_V$,
$G_V$. The observed shape of the photon spectrum unambiguously implies
a positive sign for the product $F_V G_V$. As this quantity can be related to
the low--energy constant $L_9^r(M_{\rho})$ of $\chi$PT, the
experimental results constitute an important test of the notion of
chiral vector meson dominance.

\subsection*{Acknowledgements}

We wish to thank G. Ecker for useful discussions.

\end{document}